\documentclass[reprint,superscriptaddress,amsmath,amssymb,prl]{revtex4-2}

\usepackage{amssymb}
\usepackage{amsmath}
\usepackage{physics}
\usepackage{latexsym}

\usepackage{verbatim}
\usepackage{mathrsfs}
\usepackage{amsfonts}

\usepackage{graphicx,subfigure}
\usepackage{epsfig}

\usepackage{units}
\usepackage{overpic}
\usepackage[utf8]{inputenc}
\usepackage{rotating}
\usepackage{enumerate}

\usepackage{soul}

\usepackage{xfrac}
\usepackage{xcolor}
\usepackage{multirow}

\usepackage[hidelinks]{hyperref}
\hypersetup{
    colorlinks,
    linkcolor={red!50!black},
    citecolor={blue!50!black},
    urlcolor={blue!80!black}
}

\begin{document}

\title{Thermodynamics sheds light on the nature of dark matter galactic halos}

\author{A. Ace\~na}
\affiliation{Instituto Interdisciplinario de Ciencias B\'asicas, CONICET, Facultad de Ciencias Exactas y Naturales, Universidad Nacional de Cuyo,
Mendoza, Argentina}

\author{J. Barranco}
\affiliation{Departamento de F\'isica, Divisi\'on de Ciencias e Ingenier\'ias \\
Campus Le\'on, Universidad de Guanajuato,
Le\'on 37150, M\'exico}

\author{A. Bernal}
\affiliation{Departamento de F\'isica, Divisi\'on de Ciencias e Ingenier\'ias \\
Campus Le\'on, Universidad de Guanajuato,
Le\'on 37150, M\'exico}
\affiliation{Instituto de Física, Univerisidad Michoacana de San Nicolás de Hidalgo, 58040 Morelia, Michoacán, México}

\author{E. L\'opez}
\affiliation{Observatorio Astron\'omico de Quito y Departamento de F\'isica de la Facultad de Ciencias,
Escuela Polit\'ecnica Nacional,
Quito, Ecuador}

\begin{abstract}
In spherical symmetry, the gravitational potential is uniquely determined by the rotational velocity profile. Numerous galaxies exhibit a universal velocity profile from which a universal gravitational profile is inferred. When treating dark matter as either an ideal gas, a Fermi gas, or a Bose gas, only the latter can produce a gravitational profile consistent with observations, along with a temperature profile that decreases outward. This requires the mass of the boson to be below a certain threshold of $43\mbox{ eV}/c^2$. Additionally, ensuring that the speed of sound is less than the speed of light yields a lower bound on the boson's mass.
\end{abstract}

\maketitle

{\bf \emph{Introduction:}} Current cosmological models indicate that Dark Matter (DM) is the dominant component of matter in the universe \cite{Planck:2018vyg,Fields:2006ga,WMAP:2010qai}. 
DM plays a fundamental role in explaining the kinematics of the galaxies, large-scale structure formation, and the cosmological evolution of the universe \cite{Springel:2005mi,Salucci:2018hqu}.
Its existence is inferred only through its gravitational interaction. Nevertheless, the desire to prove its existence in terrestrial experiments leads to the hypothesis that there should be an interaction of DM with baryonic matter beyond pure gravitational. 
This idea has motivated a plethora of massive particle candidates that account for DM and in addition solve some of the Standard Model of particle physics problems. To name a few: sterile neutrinos \cite{Shi:1998km}, axions and light bosons \cite{Abbott:1982af}, neutralinos, strangenets, Q-balls, weakly interacting massive particles (WIMPs) \cite{Lee:1977ua, Bertone:2004pz}, and sub-GeV massive DM particles (MeV DM).
For more than forty years, these candidates have been searched and tested with direct detection experiments \cite{PandaX-4T:2021bab,LZ:2022lsv,XENON:2023cxc}, indirect searches \cite{Barman:2022jdg,Chan:2019ptd,Song:2023xdk} and in the Large Hadron Collider \cite{CMS:2022qva,ATLAS:2022ygn}, with null discovery results. All of those experiments exclude most models that offer plausible DM candidates that interact with baryonic matter beyond gravity. Similarly, macroscopic objects such as MaCHOs \cite{EROS-2:2006ryy} and primordial black holes have been studied and ruled out as the main component of DM.

On the other hand, by neglecting DM-baryonic interactions, the so-called Cold Dark Matter (CDM) paradigm \cite{Dodelson:1996ev} has succeeded in describing DM dynamics at cosmological scales. For this model, the DM candidate is a generic nearly collisionless/pressureless particle. Through numerical simulations, CDM has reproduced the cosmic structure over an enormous span of redshifts. However, it is not possible to determine the fundamental properties of the DM from these simulations. Furthermore, such simulations predict, on small scales, an excess of substructure \cite{Weinberg:2013aya} and a theoretical density profile \cite{Navarro:1996gj} that suffers from the so-called core-cusp problem.  Another important challenge for CDM is that the Universal Rotation Curve, inferred from observations of thousands of galaxies, is in tension with the one predicted by CDM \cite{Persic:1995ru,Salucci:2018hqu}.  

In face of this crisis, a new type of investigation based on observations is needed \cite{Nesti:2023tid}. For instance, through the systematic study of stellar dynamics. There are enough high-precision data of galactic rotation curves (RC) and of dispersion velocities in clusters and elliptical/dwarf galaxies, that make galaxies an excellent natural laboratory where DM properties can be extracted. It is noticeable that the stringent bounds on a possible equation of state for DM come from analyses of RC \cite{Barranco:2013wy,Acena:2021wjx} rather than from high-precision data of cosmological surveys \cite{Muller:2004yb,Calabrese:2009zza,Xu:2013mqe,Kopp:2018zxp}.
Following this approach, in this work, properties of the DM particle are set using RC. Our hypotheses are that DM can be modeled as a fluid that forms a self-gravitating structure in hydrostatic equilibrium, which plays the role of galactic halo, and, we neglect the self-gravity of the baryonic matter, the luminous matter acts as tracer of the halo's gravitational potential. This last assumption is particularly accurate for Low Surface Brightness (LSB) galaxies, which appear to be completely dominated by DM, even in the inner regions \cite{McGaugh1998}. As DM fluids, an ideal gas, a Fermi and a Bose gas are studied. In each case, we compute the DM temperature profile of the halo consistent with RC. It is expected from the thermodynamical point of view that the temperature profile decreases as the galactic radii increase. It is found that only for the Bose gas, the temperature profile decreases outward from the galactic center and that this behavior is achieved only if the mass of the boson is below $43\mbox{ eV}/c^2$. This upper bound on the boson mass allows a good LSB RC fitting with a Burkert density profile. On the contrary, neither the ideal gas nor the Fermi gas temperature profiles decrease as the radial coordinate increases, regardless of the particle mass. We have performed the same analysis as the presented here for the Burkert density profile for different DM profiles and the results are qualitatively the same. Furthermore, if we consider the causality condition that the speed of sound has to be less than the speed of light, then for the ideal Bose gas model to be valid, the mass of the DM particle needs to be bounded below by $1.2\times 10^{-3}\mbox{ eV}/c^2$.

{\bf \emph{Velocity and temperature profiles:}}
Our starting point is the Burkert density profile for DM halos \cite{Burkert:1995yz}, that seems to correctly fit most galaxies (see \cite{Salucci:2018hqu} and references therein):
\begin{equation}
    \rho(r) = \frac{\rho_0 r_0^3}{(r+r_0)(r^2+r_0^2)},
\end{equation}
where $\rho_0$ is the central density and $r_0$ is the core radius. The DM mass inside a given radius is obtained by integrating $\frac{dM}{dr} = 4\pi r^2\rho$
and then the rotational velocity profile is $v=\sqrt{GM/r}$.
Assuming that the DM content can be described as an isotropic fluid, a gas, and that the gravitational presence of and interaction with baryonic matter can be neglected, the pressure profile, $p(r)$, is obtained by integrating the hydrostatic equilibrium equation
\begin{equation}\label{derp}
\frac{dp}{dr} = -\frac{G M \rho}{r^2},
\end{equation}
although in this case the analytical expression is not particularly enlightening.

If an equation of state is provided, $T(\rho,p)$, we immediately have the halo temperature profile, $T(r)$. Nonetheless, the thermal equilibrium equation is \cite{Hansen2004,Kippenhahn:2012qhp}
\begin{equation}\label{derT}
    \frac{dT}{dr} = -\frac{G M \rho T}{r^2 p} \nabla.
\end{equation}
The quantity $\nabla$ depends on the thermal processes that the gas undergoes. Since we do not have any {\it a priori} information on energy production and transport within the DM halo, we choose to postulate the equation of state, and then $\nabla(r)$ can be calculated from \eqref{derT}. Two particular cases stand out, the isothermal case $\nabla = 0$, and the adiabatic case $\nabla = \nabla_{ad}$, where $\nabla_{ad}$ is the adiabatic temperature gradient. This is calculated from the equation of state,
\begin{equation}
    \nabla_{ad} = \left(\frac{\partial\ln T}{\partial\ln p}\right)_s,
\end{equation}
where the subscript $s$ indicates that the derivative is calculated keeping the entropy density constant.
If for some region $\nabla(r) = \nabla_{ad}$, we speak of an "adiabatic stratification". To accept that the temperature profile obtained from the equation of state represents a state of thermal equilibrium, we impose the physical criterion that $T(r)$ has to be a non-increasing function, namely, $\nabla(r)\geq 0$ for all $r$. This criterion is consistent with the formation of a galactic halo through gravitational collapse. On the other hand, if the temperature were increasing outward, we would have to postulate an ad hoc mechanism for heat transfer against both the gravitational potential and the temperature gradient, and there is neither evidence nor theoretical justification for such a process. In fact, one of the first results in General Relativity regarding the thermodynamics of gravitating systems is the relationship between temperature and gravitational potential for a system in hydrostatic and thermodynamical equilibrium \cite{Tolman1930,EhrenfestTolman1930,Eckart1940,Klein1949}. The Tolman-Ehrenfest criterion for the thermal equilibrium of a fluid in a static spacetime, in the Newtonian approximation, states that the local thermodynamic temperature, $T$, and the Newtonian gravitational potential, $\phi$, satisfy
\begin{equation}\label{TolmanEq}
    T\left(1+\frac{\phi}{c^2}\right) = constant.
\end{equation}
This relationship has been derived using varying underlying assumptions, and even using the basic toolkit of thermodynamics, namely pistons and Carnot cycles \cite{BalazsDawson1965}, showing that a violation of \eqref{TolmanEq} implies a violation of the second law of thermodynamics. The same results have been obtained for other settings, for example, conformally static spacetimes \cite{FaraoniVanderwee2023}. With our hypothesis, and taking into account that we deal only with ideal gases, this means that the temperature has to be an outward decreasing function. Furthermore, the foundational studies on the gravothermal catastrophe show that a gravitating system of large extent cannot be isothermal \cite{LyndenBellWood1968,LyndenBellEggleton1980}. Moreover, if the initial configuration is an isothermal sphere of sufficient extent, the system is unstable and collapses by increasing without bound the central temperature. The same type of conclusion can be obtained analyzing polytropic solutions. The configurations with polytropic index below $-1$ have no boundary, the mass is unbounded and the temperature increases outward, but at the same time they are all unstable \cite{Horedt2004}. In the present setting, since we assume that DM is an ideal gas and that it is in hydrostatic equilibrium, the temperature profile has to be a decreasing function of radius for the system not to undergo a gravothermal collapse.

As we postulate the equation of state, we consider the three simplest cases:
\begin{enumerate}
\item {\it Ideal gas:} The equation of state is
\begin{equation}
    T = \frac{m}{k}\frac{p}{\rho},
\end{equation}
where $k$ is the Boltzmann constant and $m$ is the mass of the DM particle.
\item {\it Fermi gas:} The equation of state takes the parametric form
\begin{equation}
    T = \frac{m}{k}\frac{p}{\rho}\frac{f_{3/2}(z)}{f_{5/2}(z)},\quad \frac{f_{3/2}^5(z)}{f_{5/2}^3(z)} = \frac{h^6}{8\pi^3 g^2m^8}\frac{\rho^5}{p^3},
\end{equation}
being $z$ the fugacity, which is defined from the chemical potential $\mu$ through $z=e^{\mu/kT}$. Also, $h$ is the Planck constant and $g$ is the degeneracy factor, $g=2S+1$, with $S$ the spin of the particle. The functions $f_\nu(z)$ are related to the polylogarithm through $f_\nu(z) = -\mbox{Li}_\nu(-z)$.
\item {\it Bose gas:} We need to distinguish whether Bose-Einstein condensation is occurring or not. If there is no condensation,
\begin{equation}
    T = \frac{m}{k}\frac{p}{\rho}\frac{g_{3/2}(z)}{g_{5/2}(z)},\quad \frac{g_{3/2}^5(z)}{g_{5/2}^3(z)} = \frac{h^6}{8\pi^3 m^8}\frac{\rho^5}{p^3},
\end{equation}
where the functions $g_\nu$ are the polylogarithm, $g_\nu(z) = \mbox{Li}_\nu(z)$. Condensation occurs if $T<T_C$, with the critical temperature
\begin{equation}
    T_C = \frac{h^2\rho^\frac{2}{3}}{2\pi k m^\frac{5}{3} \zeta^\frac{2}{3}\left(\frac{3}{2}\right)},
\end{equation}
$\zeta$ being the Riemann zeta function, and then the temperature is
\begin{equation}\label{eqTBC}
    T = \frac{1}{k}\left(\frac{h^6p^2}{8\pi^3m^3 \zeta^2\left(\frac{5}{2}\right)}\right)^\frac{1}{5}.
\end{equation}
\end{enumerate}
For a reversible adiabatic process in an ideal Bose gas $T/p^\frac{2}{5} = \mbox{const.}$, or $\nabla_{ad}=\frac{2}{5}$, and from \eqref{eqTBC} any region of the halo where Bose-Einstein condensation occurs is automatically adiabatically stratified.

Now we turn our attention to the temperature profiles. Since the constants $\rho_0$ and $r_0$ are obtained by fitting the velocity profile of a particular galaxy, the only parameter that comes from the matter model is the mass of the particle, $m$, and for the ideal gas the temperature is proportional to it. In Figure \ref{perfilesTemp} the temperature profiles for each equation of state are shown.  These profiles correspond to the galaxy ESO3050090, which has as Burkert density profile the parameters $\rho_0 = 2.34\,(\mbox{GeV}/c^2)/\mbox{cm}^3$, $r_0 = 3.37\,\mbox{kpc}$, and for three representative values of $m$.

As stated previously, the only physical requirement that we wish to impose on the temperature profile is that it is an outward decreasing function. The temperature profile corresponding to the classical ideal gas does not satisfy this requirement, as for $r<1.12\, r_0$ it is an increasing function, and therefore we conclude that the DM halo is not composed of a classical ideal gas.

For the ideal Fermi gas, also in Figure \ref{perfilesTemp} the corresponding profiles are shown, where for simplicity we have chosen $g=2$. Analyzing the behavior of the involved functions we conclude that for the same particle mass and at the same radial position, the temperature for the Fermi gas is lower than the corresponding classical temperature. Additionally, due to the behavior of $\rho$ and $p$, for $r$ big enough the temperature in the Fermi gas is indistinguishable from the classical gas. As a function of $m$, the Fermi gas temperature is an increasing function, and for $m$ high enough the Fermi gas profile is indistinguishable from the classical gas profile. This is a consequence of the fact that in this case the classical limit of the Fermi gas is obtained for $m\rightarrow\infty$. If we lower the particle mass, at some point the temperature drops to zero. This happens first at the origin and from then on the Burkert profile can not be reproduced. This implies a minimum mass for the fermion,
\begin{equation}
    m_{min} = \left[\frac{1}{5^3}\left(\frac{3h^3}{4\pi g}\right)^2\frac{\rho_0^5}{p_0^3}\right]^\frac{1}{8},
\end{equation}
being $p_0$ the central pressure, calculated through \eqref{derp}\footnote{For a typical LSB galaxy we have $m_{min}$ around $50\,\mbox{eV}/c^2$. This is lower than the Tremaine-Gunn limit, although in the same order of magnitude. Nevertheless, we conclude that there is no fermion mass that gives a reasonable temperature profile}.
Finally, as a function of radius, there is a region starting at the origin and extending beyond the core radius in which the temperature is an increasing function, and as in the classical case we conclude that the DM halo is not composed of an ideal Fermi gas.

In the Bose case, we need to consider if there is condensation. 
Here again, the corresponding examples are shown in Figure \ref{perfilesTemp}. Contrasting with the Fermi gas, for the same $m$ and at the same $r$, the Bose temperature is higher than the classical temperature. For $r$ big enough the temperature again coincides for both profiles. As a function of $m$, the Bose gas temperature has a minimum for a certain mass, being first a decreasing function and then an increasing function of $m$. The limit $m\rightarrow\infty$ is the classical limit, therefore for $m$ big enough the Bose profile and the classical profile are indistinguishable. This means that for $m$ big enough the temperature profile is increasing in the core region. If we lower the mass, the departure from the classical behavior due to quantum effects starts to be noticeable, but still, the temperature profile has first an increasing region, a maximum, and then is decreasing. If we lower still the mass, there is a particular value, which depends only on $\rho_0$ and $r_0$, from which the temperature profile is a decreasing function for all $r$.
For this to happen, we found that typically there is first a region where the gas is found in a Bose-Einstein condensate.
As mentioned before, the region where condensation occurs is adiabatically stratified.

\begin{figure}[t]
    \centering
    \includegraphics[width=\linewidth]{proposal.eps}
    \caption{Temperature profiles for the parameters $\rho_0 = 2.34\,(\mbox{GeV}/c^2)/\mbox{cm}^3$, $r_0 = 3.37\,\mbox{kpc}$, corresponding to the galaxy ESO3050090. For the ideal Bose gas, the continuous line indicates that there is no condensation, while the dashed line indicates condensation.}
    \label{perfilesTemp}
\end{figure}

In conclusion, the only considered equation of state that implies a temperature profile that is a non-increasing function of radius is the one corresponding to a boson with low enough mass. 
In Fig. \ref{fig:bound} we present the increasing and decreasing regions of the temperature profile for two different LSB galaxies with equation of state given by the ideal Bose gas.  
Next, we explore this situation in the light of a universal relation.

{\bf \emph{Constant $\rho_{0}r_0$:}}
One particularly striking relationship among in principle unrelated quantities in galaxies is the one presented in the work \cite{Kormendy:2004se} and extended in \cite{Donato:2009ab, Gentile:2009bw,Kormendy:2014ova}. There, it is argued that the product of the central density times the core radius of the DM halo has the same value for all galaxies. The value presented is
\begin{equation}\label{valUniv}
    \rho_0 r_{0} = 141^{+82}_{-52}\,M_\odot/\mbox{pc}^2.
\end{equation}
In the present setting, this implies that the pressure profile does not depend on $\rho_0$ and only depends on $r$  through the dimensionless length $r/r_0$, or, in other words, the pressure is the same in all galactic halos modulo length rescalings. In particular, the central pressure for all halos is the same:
\begin{equation}
    p_0 = 0.83\times G(\rho_0 r_0)^2 = 4.8\times 10^{-11}\frac{\mbox{g}}{\mbox{cm}\,\mbox{s}^2}.
\end{equation}

We stick to the principle that the temperature profile has to be a decreasing function of $r$, which imposes conditions on the values of the parameters. If we take $\rho_0$ or $r_0$ as given, together with \eqref{valUniv}, then there is a maximum particle mass, $m_{max}$, above which the temperature profile fails to be an outward decreasing function. Said bound is a decreasing function of $r_0$, and we take as representative $r_0=4.82\,\mbox{kpc}$, obtained by fitting the galaxy F583, and calculate the corresponding maximum mass, which turns out to be
\begin{equation}
    m_{max} = 43\,\mbox{eV}/c^2.
\end{equation}
With these values, in each galaxy there is typically a central region where condensation occurs and which is adiabatically stratified. If we lower the mass of the DM particle or if we raise $\rho_0$, decreasing $r_0$, then the condensed region extends further away and can cover the hole observable halo of the galaxy. In Figure \ref{fig:bound} it can be seen clearly that there is a maximum particle mass for each fitted galaxy, and that the most stringent bound is provided by the more extended halo. Given \eqref{eqTBC}, the temperature profile in the central condensed region is the same for all galaxies modulo length rescalings.

\begin{figure}[t]
    \centering
    \includegraphics[width=\linewidth]{bound_mb.eps}
    \caption{Top panel: region where the temperature profile is an increasing function for the galaxies ESO3050090 and F583. The dotted lines indicate the masses used for the plots in Figure \ref{perfilesTemp}. Bottom panel: fittings of the rotation curve with the Burkert profile for said galaxies.}
    \label{fig:bound}
\end{figure}

The analysis of this section shows that not only an ideal Bose gas is compatible with \eqref{valUniv}, but that in fact it provides an explanation for such a relationship. If we take the value of $m$ to be independent of the galaxy in consideration, as it should be for the mass of a particle, then the relation \eqref{valUniv} is valid because at least the central region of all DM halos is a Bose-Einstein condensate, adiabatically stratified, and with the same central temperature:
\begin{equation}\label{TempB0}
    T_0 = \frac{1}{k}\left(\frac{h^6p_0^2}{8\pi^3m^3 \zeta^2\left(\frac{5}{2}\right)}\right)^\frac{1}{5} = \frac{0.27\,\mbox{K}}{(m[\mbox{eV}/c^2])^\frac{3}{5}}.
\end{equation}
We do not interpret \eqref{TempB0} as if all galactic halos have the exact same central temperature, but consider that there must be a fundamental reason, tied to galaxy formation and evolution, that implies that the central temperatures of galactic halos lie within a certain range. Inverting the argument, we consider the central temperature, $T_0$, as a fundamental parameter for the halo, which in turn determines the corresponding central pressure, $p_0$. As $T_0$ lies within a certain range, also $p_0$ is fairly constrained. To fully determine the structure of the halo, one needs another parameter, such as the central density. Once $T_0$ and $\rho_0$ are known, then $r_0$ can be calculated and the DM halo is determined.

The final point for consideration regards the validity of the fluid approximation, for this we take into account the thermal wavelength of the DM particle and the sound speed in the fluid. If the thermal wavelength is comparable to the size of the DM halo, then the approximation as a gas breaks down. The thermal wavelength is $\lambda = h/\sqrt{2\pi m k T}$. As a representative value, we compute this wavelength at $r=0$,
\begin{equation}
 \lambda_0 = \left(\frac{h^2\zeta\left(\frac{5}{2}\right)}{2\pi m p_0}\right)^\frac{1}{5} = \frac{1.0\times 10^{-4} \,\mbox{m}}{(m[\mbox{eV}/c^2])^\frac{1}{5}},
\end{equation}
and therefore for masses around $1\,\mbox{eV}/c^2$ the approximation of gas is perfectly valid. The thermal wavelength can be made as large as desired by reducing the mass of the particle or going far away from the center of the halo, nonetheless, even for masses far below the $\mbox{eV}/c^2$ range and in a region much bigger than the halo core the gas approximation remains valid. When the boson mass is low, the thermal de Broglie wavelength becomes large, meaning that quantum effects dominate even at moderate densities and low temperatures.
This leads to early onset of Bose-Einstein condensation in the dense central region of the halo, where gravitational confinement is strongest. The center of the halo, because of its high density, enters a partially or fully condensed state.
The condensate behaves like a macroscopically occupied quantum state that coexists with a thermal cloud.
To maintain thermodynamic equilibrium, the condensate must be in thermal contact with the outer thermal component. The central condensate region is stratified adiabatically, which means that the entropy per particle is conserved along the radius.
In systems with gravitational confinement and adiabatic stratification, the temperature decreases with increasing radius.
This behavior is analogous to the temperature gradient in self-gravitating systems such as stars, where the inner regions are hotter. When the particle mass is very low, the quantum pressure is stronger and supports a larger core against gravity.
Therefore, a larger region of the halo enters this adiabatic condensed regime with a temperature that naturally decreases outward. In a self-gravitating gas in thermal equilibrium, the Tolman relation 
($
T(r)\sqrt{g_{00}(r)} = \text{const}$) applies.
In the weak-field (Newtonian) limit, where \( g_{00}(r) \approx 1 + 2\phi(r)/c^2 \) and \( \phi(r) \) is the gravitational potential, this leads to a negative temperature gradient.
This means that temperature decreases with radius to maintain equilibrium under gravity, exactly what we observe in the case of the ideal Bose gas with condensation. On the other hand, the sound speed in the condensed region of the halo is given by
\begin{equation}
 v_s^2 = \frac{5 \zeta\left(\frac{5}{2}\right) k T}{3 \zeta\left(\frac{3}{2}\right) m} = \left(\frac{5^5 h^6 \zeta^3\left(\frac{5}{2}\right)p^2}{2^3 3^5 \pi^3 \zeta^5\left(\frac{3}{2}\right) m^8}\right)^\frac{1}{5}.
\end{equation}
If we consider the causality condition that the speed of sound has to be less than the speed of light, then for the ideal Bose gas model to be valid the mass of the DM particle needs to be bounded, $m>m_{min}$, with
\begin{equation}\label{eqMmin}
 m_{min} = \left(\frac{5^5\zeta^3\left(\frac{5}{2}\right) h^6 p_0^2}{2^3 3^5 \pi^3 \zeta^5\left(\frac{3}{2}\right) c^{10}}\right)^\frac{1}{8} = 1.2\times 10^{-3} \,\mbox{eV}/c^2.
\end{equation}
In Figure \ref{fig:prediction} we present the central temperature for the DM halos as a function of $m$, together with the excluded regions. Interestingly, the lower bound \eqref{eqMmin} excludes ultra-light axions (ULAs). This does not imply that ULAs cannot reproduce the rotation curve of galaxies, but that the ULA paradigm is incompatible with the hypothesis of the present work.

\begin{figure}
    \centering
     \includegraphics[width=\linewidth]{prediccion.eps}
    \caption{Central temperature for all DM galactic halos as function of the DM particle mass.}
    \label{fig:prediction}
\end{figure}

{\bf \emph{Conclusions:}}
We have considered three possible equations of state for the DM halos: the classical ideal gas, the ideal Fermi gas, and the ideal Bose gas. Our analysis shows that for the classical ideal gas and the ideal Fermi one, the temperature profile of the DM halo is an increasing function of radius for the core region, and therefore we deemed these cases unsuccessful at explaining the nature of DM. The ideal Bose gas produces an outward decreasing temperature profile if the mass of the particle is low enough. In light of the universal relation between central density and core radius, we conclude that the central region of the halo is in a state of Bose-Einstein condensation and that the central temperature is the same for all DM halos. Therefore, the picture of DM and the corresponding galactic halos that emerge from this work is the following. The DM particle is a boson with mass $m\leq 43\,\mbox{eV}/c^2$. The central part of the galactic halos is in a Bose-Einstein condensate with finite temperature. This condensate can extend to the whole halo. 
The condensed region is adiabatically stratified, the DM central temperature is the same for all galaxies, and the temperature profile for this region is universal modulo-length rescalings. 

We have performed the same analysis for the following profiles: PSS (Persic-Salucci-Stel), NFW (Navarro-Frenk-White), isothermal and pseudo-isothermal. The results are qualitative the same, reinforcing the paradigm presented here, and will be presented in a following paper.

\bibliographystyle{unsrt}
\bibliography{prl.bib} 

\noindent\textbf{Author Contributions} A.A. conceived and led the project. J.B., A.B., E.L. contributed to key conceptual discussions and the presentation of results. All authors contribute in the writing of the manuscript. All authors reviewed and commented on the manuscript

\noindent\textbf{Competing interests} The authors declare no competing interests.

\noindent\textbf{Correspondence and requests for materials} should be addressed to A.A. (acena.andres@conicet.gov.ar).

\end{document}